\def\unredoffs{\voffset=-1.2truecm\hoffset=-0.5truecm}
\def\redoffs{\voffset=-2.90truecm \hoffset=-1.0truecm}
\def\speclscape{}
\def\lans{l }
\ifx\answ\lans
  \speclscape
  \documentstyle[twocolumn]{article}
  \redoffs
\else
  \documentstyle[12pt]{article}
  \unredoffs
\fi
\catcode`@=11 \@addtoreset{equation}{section} \catcode`@=12

\def\yans{y }
\ifx\temp\yans
\ifx\answ\lans
\newfont{\blackb}{msbm10 scaled 1000}
\newfont{\blacks}{msbm8 scaled 1000}  
\newfont{\blackl}{msbm10 scaled 1200}  
\else
\newfont{\blackb}{msbm10 scaled 1200}
\newfont{\blacks}{msbm10 scaled 1000}  
\newfont{\blackl}{msbm10 scaled 1440}  
\fi
\else
\def\blackb{\bf}\def\blacks{\bf} \def\blackl{\bf}
\fi
\def\BC{\mbox{\blackb C}} 
\def\BR{\mbox{\blackb R}} 
 
\def\BP{\mbox{\blackb P}} 
\def\BBC{\mbox{\blacks C}} 
\def\BBR{\mbox{\blacks R}} 
 
\def\BBP{\mbox{\blacks P}} 

\def\CP{\mbox{\BC\BP}}

\def\PSL{\mbox{\rm PSL}}
\def\com#1#2{(\mbox{${}^{#1}_{#2}$})}
\begin{document}
\begin{titlepage}
\renewcommand{\thefootnote}{\fnsymbol{footnote}}
\begin{flushright}
\vbox{\begin{tabular}{c}
TIFR/TH/94-52\\
{\tt hep-th/9412078}\\
revised version\\
August 17, 1995
\end{tabular}}
\end{flushright}
\begin{center}
  {\large \bf Higher dimensional uniformisation and $W$-geometry}
\end{center}
\bigskip
\centerline{Suresh Govindarajan\footnote{E-mail: suresh@theory.tifr.res.in}}
\centerline{Theoretical Physics Group}
\centerline{Tata Institute of Fundamental Research}
\centerline{Bombay 400 005 INDIA}
\bigskip
\begin{abstract}
  We formulate the uniformisation problem underlying the geometry of
  $W_n$-gravity using the differential equation approach to
  $W$-algebras.  We construct $W_n$-space (analogous to superspace in
  supersymmetry) as an $(n-1)$ dimensional complex manifold using
  isomonodromic deformations of linear differential equations. The
  $W_n$-manifold is obtained by the quotient of a Fuchsian subgroup of
  $\PSL(n,\BR)$ which acts properly discontinuously on a simply
  connected domain in $\CP^{n-1}$. The requirement that a deformation
  be isomonodromic furnishes relations which enable one to convert
  non-linear $W$-diffeomorphisms to (linear) diffeomorphisms on the
  $W_n$-manifold.  We discuss how the Teichm\"uller spaces introduced
  by Hitchin can then be interpreted as the space of complex
  structures or the space of projective structures with real holonomy
  on the $W_n$-manifold. The projective structures are characterised
  by Halphen invariants which are appropriate generalisations of the
  Schwarzian. This construction will work for all ``generic''
  W-algebras.
\end{abstract}
\end{titlepage}
\renewcommand{\thefootnote}{\arabic{footnote}}
\setcounter{footnote}{0}

\section{Introduction}

The search for higher spin extensions of the Virasoro algebra led to
the introduction of $W_3$-algebra\cite{zam}. Unlike the Virasoro
algebra, this algebra turned out to be non-linear. $W$-algebras
associated with various simple Lie groups have been constructed. One
of the methods has been obtained from the work of Drinfeld and
Sokolov\cite{DS} who associated differential equations with Lie
Algebras and showed that the ``coefficients'' of the linear
differential equations satisfy classical $W$-algebras when equipped
with the Gelfand-Dikii Poisson bracket. For a complete list of
references, see the reviews by Bouwknegt and Schoutens, and
Feher {\it et al.}\cite{reviews}.

It is well known that the generators of the Virasoro algebra occur as
the generators of $Diff(S^1)$ as well as the residual diffeomorphisms
in the conformal gauge in two dimensional gravity (where one gets two
copies of the Virasoro algebra). It is natural to ask what sort of
structure underlies the plethora of classical $W$-algebras. It is not
clear what replaces $Diff(S^1)$\footnote{The results of this paper
  suggest that $S^1$ should also be extended to a higher dimensional
  manifold using the KdV times as extra dimensions in order to
  understand the action of W-diffeomorphisms.}. In analogy with
the Virasoro case, the generators of W-algebras should generate
residual W-diffeomorphisms of W-gravity. Extending this further, one
can construct W-string theories. W-gravity defined this way is
confined to a particular gauge and the meaning of covariance is not
very clear.  Recently, there has been some progress in this regard
using a relationship to Higgs bundles\cite{dbg,proposal}. W-gravity
can also be formulated as the theory of symmetric tensors (with the
metric tensor as the first in the series). This approach has been
pursued by Hull\cite{hull}.

In the Polyakov path integral approach to string theory, one integrates
over all metrics modulo those which are equivalent under
diffeomorphisms and Weyl transformations. On a Riemann surface
$\Sigma$ with genus $g>1$, this integral reduces to a finite
dimensional integral over the moduli space of Riemann surfaces which
is a $(3g-3)$ dimensional complex space. Related to this is the
Teichm\"uller space of Riemann surfaces which when quotiented by
``large'' diffeomorphisms (i.e., diffeomorphisms not connected to the
identity) gives the moduli space.  One expects that such a space
should exist even for the case of W-strings. In a recent
paper\cite{proposal}, it was shown that the Teichm\"uller space for
$W$-gravity is the same as one of the components of the moduli space
of certain stable Higgs bundles studied by Hitchin\cite{hitchin}.
Hitchin calls this space the {\it Teichm\"uller component}. These
spaces are a particular component of the space
Hom$(\pi_1(\Sigma);G)/G$ where $G$ is a simple Lie group.  For
$G=\PSL(n,\BR)$, we shall label the Teichm\"uller component by ${\cal
  T}_n$ with ${\cal T}_2$ being the usual Teichm\"uller space of
Riemann surfaces.

The Teichm\"uller space of Riemann surfaces, ${\cal T}_2$ is
associated with the uniformisation of Riemann surfaces. Since ${\cal
  T}_2$ is the Teichm\"uller component of Hom$(\pi_1(\Sigma);G)/G$
with $G=\PSL(2,\BR)$, every element of this space furnishes a Fuchsian
subgroup of $\PSL(2,\BR)$ which is isomorphic to the fundamental group
of $\Sigma$. This Fuchsian group acts properly discontinuously on the
half-plane with Poincar\'e metric and the Riemann surface $\Sigma$ can
be represented as the quotient of the half-plane by the Fuchsian
group. This is the uniformisation theorem for surfaces of genus $g>1$.

It is natural to ask what kind of uniformisation is associated with
the other Teichm\"uller spaces ${\cal T}_n$? This paper attempts to
answer this question. From the work of ref. \cite{proposal}, it is clear
that this question is also intimately related to the geometry of
$W$-gravity. We choose to address this question using the differential
equation approach to uniformisation. For the case of the usual
uniformisation of Riemann surfaces, this was studied by Poincar\'e and
more recently by Hejhal\cite{hejhal}. One can associate a monodromy
group $\Gamma$ to an $n$-th order Fuchsian differential
equation\footnote{On analytically continuing solutions of the
  differential equation along non-trivial loops on the surface, the
  solutions go into linear combinations of each other. So one can
  associate a monodromy matrix for every loop. The analyticity of
  solutions implies that this matrix depends only on the homotopy
  class of the loop.  The set of the monodromy matrices form the
  monodromy group $\Gamma$.}.  Generically, $\Gamma$ is a subgroup of
$\PSL(n,\BC)$.  By choosing a situation such that the monodromy group
$\Gamma$ is an element of ${\cal T}_n$, we end up having  a situation
where it is a subgroup of $\PSL(n,\BR)$.

The differential equation furnishes (locally) a map from the Riemann
surface to $\CP^{n-1}$. Lifting the differential equation to the
universal cover of the Riemann surface (which we shall model by the
half-plane with Poincar\'e metric), we obtain a map from the
half-plane to $\CP^{n-1}$. Let the image of the half-plane be
represented by $\Omega\subset\CP^{n-1}$. The monodromy group acts
properly discontinuously on $\Omega$. So we create a $W$-surface by
the quotient $\Omega/\Gamma$. There exist a set of deformations of the
differential equation which preserve $\Gamma$ -- these are the
isomonodromic deformations and they are parametrised by the times of
the generalised KdV hierarchy associated with any linear differential
operator. Labelling the first $(n-1)$ times $(t_1=z,
t_2,\ldots,t_{n-1})$, we obtain a $\Omega(\{t_i\})$ for every time
slice with the same group $\Gamma$ acting properly discontinuously on
them. We form the $(n-1)$ dimensional complex manifold
$\widetilde{\Omega}$ by the union of all the $\Omega(\{t_i\})$. We
form the $W$-manifold by the quotient $\widetilde{\Omega}/\Gamma$.
This manifold which we shall call $W_n(\Sigma)$ provides us with
$W$-space with the KdV times as the $w$-coordinates.  The uniformisation
we have just described is the main
result of this paper. A special set of diffeomorphisms on this
manifold project down to the $W$-surface as W-diffeomorphisms.

The possibility of using the KdV times to linearise diffeomorphisms is
not new\cite{GLM,gervais,gomis}.  However, what is new is the use of
the fact that they provide isomonodromic deformations. This makes the
use of KdV times more natural especially in the context of Fuchsian
uniformisation. It is natural to restrict diffeomorphisms on the
$W$-manifold to those that preserve the form of the equations
 (\ref{eexta}) which imply isomondromic deformations.  This
method has been used recently by Gomis {\it et al.}. They have
obtained finite $W$-transformations and have shown that this set
of diffeomorphisms close under composition i.e., form a
semi-group\cite{gomis}. This is indeed a remarkable result.  This
paper\cite{gomis} does have some overlap with the present work.
However, since our approaches are distinct, the results complement each
other.

There has been a large amount of work done in the context of
$W$-geometry. We shall briefly mention some of them. One of the
earliest is the work of Sotkov and Stanishkov\cite{sotkov} who
interpreted W-geometry as affine geometry.  Some interesting
observations have also been made in the Cargese lectures of
Itzykson\cite{itzykson}. Gervais and Matsuo have studied $W$-geometry
in the context of Toda theories and discuss embeddings into $\CP^n$.
They associate singularities of the embeddings with global
indices\cite{gervais}.
But their work does not seem to be related to the Fuchsian uniformisation
presented in this paper.\footnote{The difference
 can be seen by considering the well
understood case of Fuchsian uniformisation of compact Riemann surfaces.
Here the uniformisation map is not an embedding but a multi-valued
(polymorphic) map with no singularities.
Another difference, is that the natural metric related to
uniformisation is not the Fubini-Study metric considered in \cite{gervais}
but the hyperbolic Poincar\'e
metric. Thus the maps to ${\BBC\BBP}^n$
discussed in this paper are not the same as those discussed by Gervais and
Matsuo.  It would however be of interest to interpret their results
in the present context.}  Alvarez-Gaum\'e discusses the
problem of uniformisation in his ICTP lectures\cite{alvarezgaume} and
more recently, Aldrovandi and Falqui have also attempted to understand
uniformisation using period maps\cite{falqui}.

We now summarise the results of this paper.
The key result of the paper is the formulation of a Fuchsian uniformisation
related to the Teichm\"uller spaces of Hitchin. Other results are:
A direct relationship
to W-gravity is presented by relating the Beltrami differential's of the
uniformisation
to the Beltrami differentials of W-gravity (in sec. 4) and thus recovering
W-diffeomorphisms from a subgroup of diffeomorphisms on the higher dimensional
manifold.  This proves that the manifolds obtained from uniformisation can
be considered as w-space. We also discuss how higher dimensional
generalisations
of the Schwarzian, which we construct play the role of projective connections
on the higher dimensional manifold.

The paper is organised as follows. In section 2, we discuss the basics
of linear differential equations and introduce Halphen invariants
which are generalisations of the Schwarzian. We then discuss the
conditions under which a deformation of a linear differential equation
is isomonodromic. This section is mainly a collection of well known results
and are included for completeness and to fix the notation. In section 3, we
formulate the uniformisation
problem and construct $W$-space. In section 4, we discuss how
deformations of complex structure in $W$-space project to give the
standard Beltrami equations on the $W$-surface. We also show how
deformations of complex structure can be parametrised by higher
symmetric differentials. In section 5, we briefly discuss the
projective structures on $W$-manifolds and present higher dimensional
generalisations of the Schwarzian. In section 6, we end with some
conclusions and suggestions.

\section{Linear differential equations}

The relationship of linear differential equations to $W$-algebras is
quite well known. We refer the reader to the work of Drinfeld and
Sokolov\cite{DS} and Di Francesco {\it et al.}\cite{diFr} for details.
In this section, we shall introduce Halphen invariants which are
generalisations of the Schwarzian and then discuss isomonodromic
deformations of linear differential equations.
\subsection{Basics}
Consider the  homogeneous linear differential operator $L$ given by
\begin{equation}
  \hat{L} \equiv D^n + \com{n}1\hat{u}_1(z)\ D^{n-1} + \cdots +
  \hat{u}_n(z) \quad,
 \label{ediffa}
\end{equation}
where $D={d \over {dz}}$. The differential equation
\begin{equation}
\hat{L}\ \hat{f} =0\quad, \label{eodea}
\end{equation}
has locally $n$ linearly independent solutions which we shall label by
$\{f_i\}$. We shall assume that it is of the Fuchsian type, i.e., it
has only a finite number of regular singular points and no irregular
singular points. (This implies that the $\hat{u}_i$ have at most poles
of order $i$ at the singular points and are analytic elsewhere.)
In this paper, we shall be restrict ourselves to the case of no singularities.
However, we shall be solving these differential equations on Riemann
surfaces of genus larger than one and thus the non-triviality comes
from the non-trivialness of its fundamental group. Singularities correspond to
punctures (on the Riemann surface) and we shall not discuss them here.
By means of the following redefinition $ f = exp(\int \hat{u}_1)\
\hat{f}$, the differential equation can always be converted to one of
the form
\begin{equation}
L\ f = 0\quad,
  \label{eodeb}
\end{equation}
where
\begin{equation}
L \equiv D^n + \com{n}2 u_2(z)\ D^{n-2} + \cdots + u_n(z)\quad.
  \label{ediffb}
\end{equation}
The vanishing of the coefficient of $D^{n-1}$ implies that the
Wronskian is a constant (chosen to be equal to $1$ with no loss of
generality). For this to be consistent globally on the Riemann surface,
the Wronskian has to
transform like a scalar. A simple calculation (see appendix
\ref{wronk}) shows that this implies that $f$ is a ${{(1-n)}\over2}$
-- differential.  For example, for $n=2$, this differential equation
has an important role in the uniformisation of Riemann surfaces. We
obtain the well known result that $(-{1\over2})$ -- differentials play
a crucial role in the context of Riemann surfaces\cite{hawley}. The
transformation properties of the $u_i$ under holomorphic change of the
independent variable $z$ is somewhat complicated except for $u_2$
which transforms like the Schwarzian. However, there exists an
invertible (non-linear in $u_2$) change of variables $u_i\rightarrow
w_i$ where $w_i$ for $i>2$ transform as tensors of weight $i$. This
change of variables is linear when $u_2=0$ and uniquely fixed by the tensor
weight of $w_i$ and the linear dependence on $u_i$\cite{wilc,diFr}.
When $u_2=0$, one can explicitly write out the expression for $w_i$ in
terms of $u_i$ and its derivatives.
\begin{equation}
w_m = {1\over2} \sum_{s=0}^{m-3} (-)^s {{(m-2)!m!(2m-s-2)!}\over
  {(m-s-1)!(m-s)!(2m-3)!s!}} u_{m-s}^{(s)}\quad,
  \label{euwtrans}
\end{equation}
for $m=3,\ldots,n$. It is interesting to note that the above
expressions are the same for all $n$. This explains the unusual choice
normalisations in (\ref{eodeb}). The dependence on $u_2$ which is
non-linear can be restored using an algorithm described in \cite{diFr}.

One can always choose a coordinate system with $u_2=0$. The
differential equation (\ref{eodeb}) in this coordinate system is said
to be of the {\it Laguerre-Forsyth} form. This coordinate
system is obtained from the solutions of the following auxiliary
second order differential equation
\begin{equation}
(D^2 + {3\over{n+1}} u_2) \theta =0\quad.
  \label{eaux}
\end{equation}
Let $\theta_1,\theta_2$ be two linear independent solutions of this equation.
The coordinate system where $u_2=0$ corresponds to the coordinate
$w=\theta_1/\theta_2$. The residual transformations which preserve the
Laguerre-Forsyth form are the M\"obius transformations.

When all $w_i=0$ for $i>2$, all solutions of (\ref{eodeb}) can be
generated from the two solutions $\{\theta_1, \theta_2\}$ of the
auxiliary differential equation (\ref{eaux}).  The solutions are given
by $\theta_1^{n-1}$, $\theta_1^{n-2} \theta_2$, \ldots,
$\theta_2^{n-1}$. (This is the generalisation of the $1$, $z$, \ldots,
$z^{n-1}$ when all $u_i=0$.) Further, all the $u_i$ for $i>2$ are
completely determined in terms of $u_2$ and its derivatives. For
example, $u_3 = {3\over 2} u'_2$ and $u_4 ={9 \over 5} u_2'' +
{{3(5n+7)} \over {5(n+1)}} u_2^2$\cite{forsyth}.

For the case of $n=3$, it is easy to see that the solutions of the
differential equation satisfy the quadratic  relation
\begin{equation}
f_1 f_3 - (f_2)^{2}=0\quad.
  \label{ecan}
\end{equation}
The subgroup of $\PSL(3,\BC)$ that preserves the relation is $\PSL(2,\BC)$.

\subsection{Schwarzian and it generalisations}

The $\{f_i\}$ can be thought of (locally) as homogeneous coordinates
on $\CP^{n-1}$. We can define inhomogeneous coordinates $s_i$ for
$i=1,\ldots,(n-1)$ by $s_i = (f_{i}/f_n)$. The $\{u_i\}$ now play the
role of {\it projective invariants} which characterise the map
into $\CP^{n-1}$. Projective invariants are differential
functions $I^r(\{s_i\};z)$ which are invariant under the group of
projective transformations. For example, for the case of $n=2$, the
Schwarzian is the invariant.  We shall now show how one constructs
these invariants. We shall follow the construction described in
Forsyth\cite{forsyth} and Wilczynski\cite{wilc}.  In his study of
plane and space curves (n=3,4 in our language), Halphen constructed
quantities from the inhomogeneous coordinates which were invariant
under projective transformations. Hence the $I^r$ are called the {\it
  Halphen invariants} by Wilczynski. Painlev\'e and Boulanger also
derived the invariants for $n=3$. These have
been derived more than a century ago in the study of classical
invariant theory.

The Schwarzian will be denoted by $S(s;z)$. It has the
following properties
\begin{eqnarray}
  S({{as+b}\over{cs+d}};z) = S(s;z)\quad,\nonumber{}\\
  S(s;z) = S(s;t)({{dt}\over{dz}})^2 + S(t;z)\quad,
\label{epropa}
\end{eqnarray}
where the first equation is the statement of projective invariance of
the Schwarzian and the second gives its transformation under
holomorphic change of coordinates. Under M\"obius transformations, it
shows that the Schwarzian behaves as a modular form of degree two.

The projective invariants $I^r$ for $r=2,\ldots,n$
have the following properties (Compare with (\ref{epropa}).)
\begin{eqnarray}
  I^r(A \vec{s};z)=I^r(s;z)\nonumber{}\quad,\\
  I^2(\vec{s};z) = I^2(\vec{s};t)({{dt}\over{dz}})^2 + {{n+1}\over6}
  S(t;z)\quad,\nonumber{}\\
  I^r(\vec{s};z) = I^r(\vec{s};t)({{dt}\over{dz}})^r\quad,\ n>3
  \label{epropb}
\end{eqnarray}
where the superscript $r$ indicates the weight of the invariant under
transformations of the independent variable and $A\in\PSL(n,\BC)$. It
will be shown that solutions of $n-th$ order differential equation
(\ref{eodeb}) give rise to solutions of the following coupled set of
non-linear differential equations
\begin{equation}
I^r(\vec{s};z)=w_r(z)\quad.
  \label{enona}
\end{equation}
The invariance under projective transformations, $\vec{s}\rightarrow A
\vec{s}$, can be seen from the projective invariance of the right hand
side of the above equation.  For the case $n=2$ this gives the well
known result that solutions of the non-linear differential equation
$S(f;z)=u_2(z)$ can be constructed from the ratio of any two solutions
of the second order differential equation (\ref{eodeb}).

The differential invariants are constructed as follows\cite{forsyth}.
We shall explicitly illustrate the steps using the $n=2$ case.
Substitute $f_i=s_{i}f_n$ for $i=1,\ldots,(n-1)$ in the differential
equation (\ref{eodeb}). We obtain $(n-1)$ equations (call them Set A)
linear and homogeneous in $f_n$ and its derivatives.
$$
{\rm Set A:}\qquad\qquad\quad s^{''}f_2 +2 s^{'}f_2^{'} =0\quad\qquad.
$$
By differentiating these equations, we obtain
$(n-1)$ equations (call them Set B) linear and homogeneous in $f_n$
and its first $(n-1)$ derivatives.
$$
{\rm Set B:}\quad\quad(s^{'''}-2s^{'}u_2)f_2 + 3 s^{''}f_2^{'}=0\quad.
$$
Choose one equation from Set B and all of Set A to obtain $n$ linear
equations in $f_n$ and its derivatives. The existence of non-trivial
solutions $f_n$ implies that the determinant of the matrix of the
coefficients of $f_n$ and its derivatives vanish. We thus obtain
$(n-1)$ equations, one for each equation in Set B.
$$
\left|\matrix{ s^{''} &2 s^{'}\cr (s^{'''}-2s^{'}u_2) & 3 s^{''}}\right|=0
$$
Solving these $(n-1)$ equations for the variables $u_i$, we obtain
expressions for the $u_i\equiv J^i(\vec{s};z)$ in terms of the
inhomogeneous variables and their derivatives. Making the change of
variables $u_i\rightarrow w_i$, we obtain the expressions
$w_r=I^r(\vec{s};z)$.  As is clear, the $n=2$ case gives the
Schwarzian. Explicit expressions for $n=3$ are given in Forsyth's
book\cite[see page 195]{forsyth}.

{\it There exists a bijective map from the set of solutions of the
linear homogeneous differential equation (\ref{eodeb}) and the
non-linear differential equation (\ref{enona}).}\\ The proof of this
is a follows. We have just seen, $s_i=f_{i}/f_n$ for
$i=1,\ldots,(n-1)$ gives us solutions $\vec{s}$ of (\ref{enona})
provided $f_i$ solve (\ref{eodeb}). To prove the converse, one
can show that
\begin{equation}
J^{-{1\over n}}\equiv f_n = \left|\matrix{s_1^{'}&\cdots&s_1^{(n-1)}\cr
                    \vdots  & \ddots & \vdots \cr
                   s_{n-1}^{'}&\cdots&s_{n-1}^{(n-1)}}\right|^{-{1\over n}}
\label{efone}
\end{equation}
and $f_{i}=s_i f_n$ are solutions of (\ref{eodeb}) provided
${\vec{s}}$ solve (\ref{enona}). The expression
for $f_n$ follows from property (\ref{ewprop}) of the Wronskian.
Choose $\lambda=f_n$ and using the fact that the Wronskian is a
constant (chosen to be $1$), we recover the expression for $f_n$.
Hence the solutions of (\ref{eodeb}) can be written in terms of
$\vec{s}$ as $f_{i}=s_i f_n$ for $i=1,\ldots,(n-1)$ and $f_n$ as
given above.

\subsection{Isomonodromic Deformations and the generalised KdV
 hierarchy}

Let $\{f_i\}$ for $i=1,\ldots,n$ be a basis of linearly independent
solutions of the differential equation (\ref{eodeb}) with Wronskian
equal to 1. Let $\gamma$ be a closed path (possibly around a regular
singular point or a non-trivial cycle with base point $z_0$).  On
analytically continuing the solutions $f_i$ around the path $\gamma$,
we obtain another basis $\{\check{f}_i\}$. Since there are only $n$
independent solutions, the new basis $\{f'_i\}$ is related linearly to
the old basis
$$
\check{f}_i = [M_\gamma(z_0)]_i^{~j}\ f_j\quad,
$$ where $M_\gamma(z_0)$ is a constant (independent of $z$)
PGL$(n,\BC)$ matrix which becomes a $\PSL(n,\BC)$ matrix due to the
condition on the Wronskian. The analyticity of the solutions in the
neighbourhood of non-singular points implies that $M_\gamma(z_0)$
depends only on the homotopy class of $\gamma$ and the choice of basis
$\{f_i\}$.  $M_\gamma$ are referred to as {\it monodromy matrices}.
The monodromy matrices furnish a map from $\pi_1(\Sigma\backslash E)$
to $\PSL(n,\BC)$, where $E$ denotes the set of regular singular points
of (\ref{eodeb}). The set of matrices $M_\gamma$ form a group which we
shall call the {\it monodromy group}. Further, the freedom in the
choice of basis implies that this map is well defined up to an overall
conjugation in $\PSL(n,\BC)$. Hence the monodromy of a Fuchsian
differential equation furnishes an element of Hom$(\pi_1;\PSL(n,\BC))$
well defined up to overall conjugation in $\PSL(n,\BC)$. We shall
refer to the conjugacy class as the {\it monodromy homomorphism} of
the differential equation or simply the monodromy homomorphism.

Let $T_\gamma(z_0)$ be the ``generator'' of translation along
$\gamma$. So we have
\begin{equation}
T_\gamma(z_0) \vec{f} = M_\gamma(z_0) \vec{f}\quad.
  \label{emonod}
\end{equation}
Deformations of equation (\ref{eodeb}) such that the monodromy group
is preserved are called {\it isomonodromic deformations}. We shall
show that flows of the generalised KdV hierarchy, such that
(\ref{eexta}) is satisfied, generate such deformations. Isomonodromic
deformations have been studied in the context of linear differential
equations with regular as well as irregular singular points on $\CP^1$
in \cite{jimbo, its}. We shall however base our discussion on the work
of Novikov where he discusses changes of the monodromy matrix under
deformations in the context of the second order (Sturm-Liouville)
equation with periodic potential\cite{novikov}.

The generalised KdV hierarchy associated with the operator $L$ is
defined by the following set of non-linear differential equations in
$u_i$\cite{DS}.
\begin{equation}
{d\over {dt_p}} L = [(L^{p/n})_+,L]\quad,
  \label{ekdv}
\end{equation}
where $p=1,2,\ldots$ and by $(L^{p/n})_+$, we mean the differential
operator part of the pseudo-differential operator $(L^{p/n})$. Here
the $u_i$ are extended as functions of $t_p$. For example, in the $n=3$
case, the $u_2$ and $u_3$ satisfy the equations of the Boussinesq
hierarchy. The variables $t_p$ represent the ``times'' of the
generalised KdV hierarchy. We will be consider only the first $(n-1)$
flows and one can choose $t_1=z$.

Taking the derivative of the differential equation $L\vec{f}=0$ with
respect to the time variable $t_p$ and
substituting the flow condition (\ref{ekdv}), we get
\begin{equation}
L\left(\partial_p \vec{f} - (L^{p/n})_+ \vec{f}\right) =0\quad.
  \label{emonoda}
\end{equation}
Note that we have extended $\vec{f}$ as functions of all the KdV
times.  This implies that the term in the brackets above also solves
the differential equation (\ref{eodeb}). Since $\vec{f}$ give a basis
of solutions, we can write
\begin{equation}
(\partial_p \vec{f} - (L^{p/n})_+ \vec{f})= \Lambda_p \vec{f}\quad,
  \label{emonodrela}
\end{equation}
where $\Lambda_p$ is some ($z_0$ and $t_p$ dependent) matrix. In order
to obtain the time dependence of the monodromy matrices
$M_\gamma(z_0)$, we will apply the translation operation
$T_\gamma(z_0)$ on equation (\ref{emonodrela}) to obtain the following
equation,
\begin{equation}
\partial_p M_\gamma(z_0) = [\Lambda_p,M_\gamma(z_0)]\quad,
  \label{emonodrelb}
\end{equation}
where $\Lambda_p$ does not depend on the choice of $\gamma$.
Since, $M$ is an element of the Lie group $\PSL(n,\BC)$ for all $t_p$,
it follows that $\Lambda_p$ is an element of the corresponding Lie
Algebra. Further, integrating the equation, we obtain
\begin{equation}
M_\gamma(t_p)= g(t_p)\  M_\gamma(0)\  g^{-1}(t_p)\quad,
  \label{emonodrelc}
\end{equation}
where $g(t_p)=P(exp\int_0^{t_p} dt_p \Lambda_p)$ is an element of
$\PSL(n,\BC)$. Hence, the conjugacy class of the monodromy homomorphism
remains invariant for an arbitrary KdV flow.

When $\Lambda_p$ is proportional to the identity matrix, we obtain from
(\ref{emonodrelb}) that the monodromy matrices are time independent.
This implies that such KdV flows are isomonodromic. Choosing this
constant to be zero and substituting in equation (\ref{emonodrela}),
we get
\begin{equation}
\partial_p \vec{f} = (L^{p/n})_+ \vec{f}  \quad.
\label{eexta}
\end{equation}
Hence, the generalised KdV flows generate isomonodromic deformations
when (\ref{eexta}) is satisfied. This is probably a well known result
but we are unaware of a reference providing this result in the context
of linear differential equations on a Riemann surface.

\section{Higher dimensional uniformisation}

The solution of second order Fuchsian differential equations on
Riemann surfaces of genus $g>1$ are related to the uniformisation of
Riemann surfaces. This was studied extensively by Poincar\'e and more
recently by Hejhal\cite{hejhal}. This has not been a very successful
approach to uniformisation partly due to the problem of the so called
accessory parameters.  The higher order Fuchsian differential
equations should also be related to what we shall call {\it higher
  dimensional uniformisation}.  Just as in the case of $n=2$, this
approach will continue to be plagued by accessory parameters.
Nevertheless, inspite of this problem, this does provide a better
understanding of the meaning  of higher dimensional
  uniformisation. The Teichm\"uller space related to this
uniformisation will be the Teichm\"uller components of Hitchin, which
we shall denote by ${\cal T}_n$.

In the usual uniformisation of Riemann surfaces, for the case of genus
$g>1$, the uniformisation theorem tells us that the Riemann surface
$\Sigma$ can be represented by the quotient of the half plane (which
is a domain in $\CP^1$) by the action of a Fuchsian subgroup of
$\PSL(2,\BR)$.  This Fuchsian subgroup is isomorphic to the
fundamental group of the Riemann surface. Each point in Teichm\"uller
space corresponds to a different representation of the fundamental
group as a Fuchsian group.  Generalising this, we expect a surface
$\Omega$ embedded in $CP^{n-1}$ on which a Fuchsian subgroup $\Gamma$
of $\PSL(n,\BR)$ which is isomorphic to the the fundamental group of
the Riemann surface\footnote{Since there does not seem to be a general
  definition of a Fuchsian subgroup of $\PSL(n,\BBR)$, we shall require
  that it be a subgroup of $\PSL(n,\BBR)$ which acts properly
  discontinuously on $\Omega$ with no fixed points. It may be
  necessary to impose more conditions in the definition of a Fuchsian
  group but we shall ignore these technicalities for now.}.  One then
obtains a $W$-surface which is given by $\Omega/\Gamma$. We will show
that there is an extension of this surface to a $(n-1)$ dimensional
manifold obtained by isomonodromic deformations. This manifold
provides us with W-space.  This is the primary result of this paper.

The solutions $\{f_i\}$ of (\ref{eodeb}) furnish (polymorphic) maps
from the Riemann surface $\Sigma$ to $\CP^{n-1}$ where $f_i$ are the
homogeneous coordinates.  Choosing coordinates such that the
differential equation has the Laguerre-Forsyth form has a nice
interpretation here. This coordinate can be identified with the
coordinate of a fundamental domain in the half-plane with Poincar\'e
metric. We shall refer to the map from the half-plane to $\CP^{n-1}$
as the {\it developing map}.

The developing map furnishes us with a domain $\Omega \subset
\CP^{n-1}$ as the image of the half-plane. However, the monodromy
homomorphism generically gives a subgroup of $\PSL(n,\BC)$ which is
not exactly what one wants.  Nevertheless, the differential equation
approach does provide a concrete realisation of how a higher
projective structure may be realised on Riemann surfaces (see section
5). Also, for the purposes of our discussion we shall assume that the
monodromy homomorphism gives a subgroup $\Gamma$ of $\PSL(n,\BR)$
which corresponds to a point in the Teichm\"uller component ${\cal
  T}_n$. In other words, we shall assume that the differential
equation being considered is the one related to the uniformisation
problem we are formulating here.

\subsection{W-space from isomonodromic deformations}

Let $\tilde{f}_i(t_p)$ be extensions of the solutions $f_i$ of the
(\ref{eodeb}) such that deformations parametrised by $t_p$ are
isomonodromic. As we have shown, this requires that $\tilde{f}$
satisfy equations (\ref{eexta}).  Further, we impose the boundary
condition $\tilde{f}(t_i)=f(z)$ at $t_p=0$ for $i>1$.

The differential equation gives maps into $\CP^{n-1}$ for every time
slice labelled by $\{t_i\}$. The corresponding developing map provides
us with $\Omega(t_i)$.  The same Fuchsian subgroup $\Gamma$ of
$\PSL(n,\BR)$ acts on $\Omega(t_i)$ since the deformations are
isomonodromic by choice. We can now form the $(n-1)$ dimensional
complex manifold\footnote{One has to show that $\tilde{\Omega}$ forms
  a manifold. As we shall see later, for the case where all $u_i$
  vanish, it is easy to see that the $\widetilde{\Omega}$ is a
  manifold.  This will continue to be true for the case $u_i\neq0$
  since diffeomorphisms will not change this.}
$\widetilde{\Omega}\equiv \cup \Omega(t_i) \subset \CP^{n-1}$.
$\Gamma$ continues to act discontinuously on the whole of
$\widetilde\Omega$. We then obtain a {\it W-manifold} $W_n(\Sigma)$ of
dimension $(n-1)$ by $\widetilde\Omega /\Gamma$. Hereafter, we shall
refer to the space $\widetilde\Omega/\Gamma$ as the $W$-manifold and
$\Omega/\Gamma$ as the $W$-surface. It is clear that
$\widetilde{\Omega}$ plays the role of the half-plane in the standard
uniformisation of Riemann surfaces. As we shall demonstrate later,
this manifold provides the setting for the linearisation of
$W$-diffeomorphisms.  The extended homogeneous coordinates
$\tilde{f}_i$ are related to the the Baker-Akhiezer function with only
a finite number of non-zero times (see the work of Matsuo for more
details\cite{matsuo}).

It is of interest to study the case where all the $u_i$ vanish.  The
$\tilde{s}_i$ are related to the $t_i$ by the relation $\tilde{s}_i =
{{z^i}\over{i!}} + t_i+\cdots$ and $\tilde{J}=1$. The ellipsis refers
to terms involving ``times'' $t_j$ for $j<i$. From this it is clear that
the $W$-surface at time $t_i$ has no overlap with the $W$-surface at a
different time slice.  This implies that $\Omega(t_i) \cap
\Omega(t'_i)=0$ for $t_i\neq t'_i$.  We expect that this situation
will continue to hold for the case $u_i\neq0$. We shall call the
$W$-manifold in this case the {\it trivial} W-manifold. For example, for the
case $n=3$, the quadratic relation (\ref{ecan}) gets modified to
\begin{equation}
f_1f_3 - f_2^2 +2t f_1^2 =0\quad.
\end{equation}
Since $\Omega(t_i)$ are simply connected and are time translations of
$\Omega(0)$, $\widetilde\Omega$ which is the union of all the
$\Omega(t_i)$ is also simply connected.  Hence the fundamental group
of the {\it W-manifold} is the same as that of the original Riemann
surface.  This also suggests that the directions corresponding to
$t_i$ for $i>2$ are probably non-compact.

The $W$-manifold is a hyperbolic manifold (with a non-trivial
Kobayashi pseudo-metric induced from the developing
map)\cite{kobayashi}.  It is of interest to see if the pseudo-metric
is a metric. This metric however cannot be the Fubini-Study metric. It
is possibly a Kahler manifold but we do not know how to establish
this.  The $W$-manifold described here does not seem to have any relation
to the manifold introduced by Zucchini as a possible candidate for
$W$-space\cite{zucchini}.

Let us define extended inhomogeneous coordinates by $\tilde{s}_i =
\tilde{f}_{i}/\tilde{f}_n$ where $\tilde{f_i}$ are the extensions of
$f_i$. Interpreting $\tilde{s}_i$ as an arbitrary holomorphic
coordinate transformation of $t_i$, eqn.  (\ref{efone}) has a nice
interpretation -- it is the $-{1\over n}$-th power of the determinant
of the Jacobian of a holomorphic coordinate transformation. This
follows from using eqn. (\ref{eexta}) to replace higher $z$
derivatives of $\tilde{s}_i$ by first order derivatives in $t_i$ and
some standard properties of determinants.
\begin{equation}
\tilde{J}^{-{1\over n}}\equiv \tilde{f}_n = \left|\matrix{\partial_1
  \tilde{s}_1&\cdots&\partial_{n-1} \tilde{s}_1\cr
                    \vdots  & \ddots & \vdots \cr
\partial_1  \tilde{s}_{n-1}&\cdots&\partial_{n-1} \tilde{s}_{n-1}}
\right|^{-{1\over n}}\quad,
\label{eftwo}
\end{equation}
where $\partial_i=\partial/\partial t_i$ and $\tilde{s}$ are the
extended inhomogeneous coordinates. This implies that the homogeneous
coordinates transform like tensor densities of weight $-{1\over n}$ on
the $W$-manifold.

On the Riemann surface, $J$ transforms as a tensor density of weight
$n(n-1)/2$, i.e., it is a section of $K^{{n(n-1)}\over2}$. Similarly,
in the higher dimensional case, $\tilde{J}$ can be considered as a
section of the determinant bundle of the canonical bundle on the
$W$-manifold.  So a field of weight $\Delta$ on the Riemann surface
which is a section of $K^\Delta$ can be lifted to the higher
dimensional case as a section of $[2\Delta/(n^2-n)]$-th power of the
determinant bundle. This observation should enable us to derive the
transformation of arbitrary tensors under W-diffeomorphisms.

In $W$-space not all diffeomorphisms are allowed. The diffeomorphisms
which preserve the form of (\ref{eodeb}) and (\ref{eexta}) will form
the set of restricted diffeomorphisms. Since this has already been
recently studied by Gomis {\it et al.} and we refer the reader to
their work\cite{gomis}. We shall obtain the conditions using a
slightly different method in the next section. Gomis {\it et al.} have
also shown that these restricted diffeomorphisms form a semi-group i.e.,
the composition of two restricted diffeomorphisms gives an restricted
diffeomorphism.  $W$-diffeomorphisms are then obtained by projecting
the restricted diffeomorphisms to the $W$-surface.

\section{Deformation of complex structure}

In this section, we shall first introduce the Beltrami equation in the higher
dimensional setting. This encodes the change of complex structure on
W-manifolds. The compatibility of this equation with the equations
of the KdV hierarchy as well as the differential equation provides
restrictions on the allowed Beltrami differentials. In order to relate
the W-manifolds to W-gravity, we then show that the higher dimensional
Beltrami equation projects to the W-surface as the Beltrami
equation which occurs in W-gravity\cite{GLM,flat}. In order to make this
correspondence we discuss an old puzzle raised by Di Francesco {\it et al.}
regarding the relationship of KdV flows and W-diffeomorphism which
appears in our setting as the relationship of the Beltrami differentials
of higher dimensional uniformisation and the Beltrami differentials
of W-gravity. We suggest a resolution to this puzzle and verify it
explicitly for the case of $W_3$ and $W_4$.

So far we have used the differential equation approach to describe the
geometry of $W$-space. However, for describing Teichm\"uller space, it
better to use other techniques like generalisations of quasi-conformal
mappings. Related to this is the Beltrami equation which describes the
deformation of complex structure. In this section, we shall discuss
how the Beltrami equation on the $W$-manifold projects down to the
$W$-surface as a non-linear Beltrami equation. Unlike the case of
isomonodromic deformations, a deformation of complex structure leads
to a deformation in the Fuchsian group and hence corresponds to a
different point in the Teichm\"uller component ${\cal T}_n$.  So far
we have been implicitly assuming that the $\tilde{f}_i$ are
holomorphic functions of $t_i$ i.e.,
\begin{equation}
\bar{\partial}_i \tilde{f} =0\quad,
  \label{ehol}
\end{equation}
where $\bar{\partial}_i = \partial / \partial \bar{t}_i$. It is
convenient to replace the above holomorphicity condition in terms of
the inhomogeneous coordinates since the usual Beltrami equation is
written in terms of these coordinates.
\begin{equation}
\bar{\partial}_i \tilde{s} =0\quad.
  \label{eholb}
\end{equation}
A deformation of complex structure can be parametrised by means of the
Beltrami differentials defined by the following equations
\begin{equation}
[\bar{\partial}_i + \mu_i^{\ j} \partial_j] \tilde{s} =0\quad.
  \label{ebeltrami}
\end{equation}
{}From this equation, we can show that
$\tilde{J}$ satisfies the following equation
\begin{equation}
[\bar{\partial}_i + \mu_i^{\ j} \partial_j + (\partial_j\mu_i^{\ j})
] \tilde{J} =0\quad.
  \label{ejacbeltrami}
\end{equation}
We shall only consider a restricted class of Beltrami deformations
given by $\mu_i^{\ j}=0$ for $i>1$. This implies that one has $(n-1)$
non-zero components corresponding to $\mu_{\bar z}^{\ j}$. We do not have any
geometric insight into why the other $\mu_i^{\ j}$ vanish but impose it
using the standard counting and conformal dimensions of the Beltrami
differentials which occur in $W$-gravity. However it
suggests the possibility that the $W$-manifold is such that it is
``rigid'' to those complex deformations parametrised by $\mu_i^{\ j}$
for $i>1$. Using $\tilde{f}_n=\tilde{J}^{-1/n}$, we obtain the
equation
\begin{equation}
  [\bar{\partial}_i + \mu_i^{\ j} \partial_j -{1\over n}
  (\partial_j\mu_i^{\ j}) ] \tilde{f} =0\quad.
  \label{ebeltramid}
\end{equation}
The compatibility of eqn. (\ref{ebeltramid}) with eqn. (\ref{eexta})
imposes restrictions on the Beltrami differentials. These are related
to the restricted diffeomorphisms obtained in ref. \cite{gomis}. These
conditions are given by the following set of equations
\begin{equation}
[\partial_p - (L^{p/n})_+\ ,\  \bar\partial_i + \mu_i^{\ j}\partial_j
-{1\over n} (\partial_j\mu_i^{\ j})]\tilde{f}=0\quad.
  \label{eresta}
\end{equation}
For $n=3$, these conditions are
\begin{eqnarray}
\partial_t\mu &=& -{1\over6}\partial_z^3\rho -2 u_2
\partial_z\rho\quad,\nonumber{}\\
\partial_t \rho &=& 2 \partial_z\mu \quad,
  \label{edifbel}
\end{eqnarray}
where $t\equiv t_2$; $\mu_{\bar z}^{\ z}\equiv (\mu - {1\over2}\partial_z\rho)$
and $\mu_{\bar z}^{\ t}\equiv \rho$. These relations are consistent
with the conditions on infinitesimal $W$-diffeomorphisms obtained in
ref. \cite{gomis}.  The Beltrami equations become
\begin{eqnarray}
[\partial_{\bar z} + (\mu -{1\over2}\partial_z\rho)\partial_z +
\rho\partial_t]\tilde{s}=0 \quad,\nonumber{}\\{}
[\partial_{\bar z} + (\mu -{1\over2}\partial_z\rho)\partial_z +
\rho\partial_t -{1\over3}\partial_z(\mu -{1\over2}\partial_z\rho)
-{1\over3} (\partial_t\rho)] \tilde{f}=0\quad.
  \end{eqnarray}
Thus equation (\ref{ebeltramid}) and conditions (\ref{edifbel}) combine
to give us the Beltrami equation on the W-manifold.

Projecting to the $W$-surface, the Beltrami equation becomes a non-linear
equation. We obtain
\begin{eqnarray}
  [\bar{\partial}_{\bar z} + (\mu-{1\over2}\partial_z\rho) \partial_z
  + \rho(\partial_z^2 -{2\over{3\tilde{J}}}\partial_z)] \tilde{s}
  =0\quad, \nonumber{}\\ {}
  [\bar{\partial}_{\bar z} + \mu \partial_z -(\partial_z\mu) +
  \rho(\partial_z^2 +2u_2)-{1\over2}(\partial_z\rho)\partial_z
  +{1\over6}(\partial_z^2\rho)] \tilde{f} =0\quad,
\label{ebeltramib}
\end{eqnarray}
where we have used $\partial_t \tilde{f} = (\partial_z^2 +2u_2)
\tilde{f}$.  This is precisely the form of the Beltrami equations
obtained in earlier work\cite{GLM,flat}.

We would like to comment on the observation by the authors of
\cite{diFr} that the generators of $W$-transformations ($w_i$ in our
notation) do not in general coincide with the generators of the
generalised KdV flows (these correspond to Res($L^{p/n}$)). This is
reflected by the fact that the Beltrami differentials corresponding
to $\mu_1^{\ i}$ are not directly the Beltrami differentials in the
usual formulation of $W$-gravity.  Nevertheless, we claim that
there exists an invertible transformation (possibly non-linear)
relating the two sets of Beltrami differentials. This transformation
corresponds to converting the $\mu_1^{\ i}$ to $(-i,1)$ tensors (i.e.,
sections of $K^{-i}{\bar K}$) on projecting to the $W$-surface.  In
the example we considered $\mu$ and $\rho$ are $(-1,1)$ and $(-2,1)$
tensors. This is the dual of the $u_i\rightarrow w_i$ transformation.
It is easy to see that the transformation of $\mu_1^{\ i}$ will
involve derivatives of all the $\mu_1^{\ j}$ with $j>i$. It is likely
that some of the $u_i$ (atleast $u_2$) will enter this transformation.
This possibility first occurs in the case of $W_4$ where one could
have a term of the type $\mu=\mu_1^{\ z} + (a \partial_z^2 +b u_2)
\mu_1^{\ t_3} + \cdots$ where $a$ and $b$ are some constants. This
explains why $Res(L^{3/4})\neq w_4$. As a check, we set $u_2=0$ in the
$W_4$ transformations given in \cite{diFr} and find that for constant
transformations that $X_3=Y_3$ (using the notation of \cite{diFr}).  It
is therefore useful to work out an algorithm to generate this
transformation. We hope to address this issue in the future\cite{beltrami}.

The Beltrami differentials which parametrise the deformation of
complex structure can be constructed using quadratic, cubic and higher
order symmetric differentials\cite{proposal}. Thus, if we identify the
space of complex structures as being locally parametrised by these
differentials, we immediately recover the result that the dimension of
this space is given by dim(PSL($n,\BR))(2g-2)=(2g-2)(n^2-1)$. This
space can be identified with the Teichmuller component of Hitchin
labelled ${\cal T}_n$. This follows by using the result that the
Fuchsian group gets deformed by the Beltrami differentials and the
space of Fuchsian groups which uniformise is the Teichm\"uller component
${\cal T}_n$.

\section{Projective structures with real holonomy}

It has been known for a while now that the generators of $W$-algebras
behave as generalised projective connections. However, the exact sense
in which this occurs was not clear. The differential equation
constructed using the projective connections did give maps into
$\CP^{n-1}$, but the fact that the map is from a one dimensional space
to a $(n-1)$ dimensional space did not let one use the standard
definition of projective structures except when $n=2$. However, we
have now been able to extend the base space to an $(n-1)$ dimensional
space and now the standard definition can be applied.

A projective structure can now be defined as maps from charts on
$W_n(\Sigma)$ to $\CP^{n-1}$ with transition functions at the overlap
of charts belonging to $\PSL(n,\BR)$. Note that the most general
transition function would belong to $\PSL(n,\BC)$ but we are
interested in real projective structures or to be precise, $\CP^{n-1}$
structures with real holonomy. This imposes the reality condition on
the transition functions. Given a patch definition of the projective
structure, one can easily obtain the corresponding projective
connections either by using the Wronskian method (if the projective
structure is described in terms of homogeneous coordinates) or the
Halphen invariants (if the projective structure is defined in terms of
inhomogeneous coordinates).

The converse is a little bit more involved. Given the projective
connections $u_i$, one would like to find the projective coordinates.
First, one has to extend the $u_i$ as functions of the KdV times such
that they solve the relevant equations of the generalised KdV
hierarchy.  Then the solutions of the equations (\ref{eodeb}) and
(\ref{eexta}) furnish the projective coordinates $f_i$. Note that one
has implicitly extended to the Riemann surface to the trivial $W$-manifold.

The Halphen invariants $I^r$ can be also be rewritten in terms of the
extended variables. For example, for the case $n=3$ we obtain
\begin{eqnarray}
I^2(s_1,s_2;z,t) &=& {1\over2} \tilde{J}^{1\over3}(\partial_t - \partial_z^2)
       \tilde{J}^{-{1\over3}}\quad,\nonumber\\
I^3(s_1,s_2;z,t)&=& {1\over2} \tilde{J}^{1\over3} (\partial_z^3 - 3 \partial_t
\partial_z)
       \tilde{J}^{-{1\over3}} + {3\over2} \partial_z (I^2(s_1,s_2;z,t))\quad.
\nonumber
\end{eqnarray}
In this higher dimensional setting, the Halphen invariants now are
the right generalisations of the Schwarzian.\footnote{We would like
to observe that the occurance of the Schur polynomials (in derivatives)
 in the expressions for the Halphen invariants. Thus, the Schur polynomials
might be useful to generate expressions for the Halphen invariants
in the general case.} Unlike the situation in section 2, now
the symmetry between the inhomogenous coordinates and
the independent variables ($z,t,\ldots$) has been restored in the $I^r$.
In the form given above, Gomis {\it et al.} refer to these as
$W$-Schwarzians but we shall continue to call them Halphen invariants.
It is actually simpler to use the equations (\ref{eexta}) to obtain
the extended Halphen invariants by substituting $\tilde{J}^{-{1\over
    n}}$ for $f$ and solving for $u_i$.  The Halphen invariants
continue to be invariant under projective transformations even after
they are extended by the KdV times.  Further, they vanish when the
coordinates $t_i$ are chosen such that $s_i$ are related to the $t_i$
as in the trivial $W$-manifold. In this coordinate system,
$\tilde{J}=1$ and by recursively using the general structure of
equations in (\ref{eexta}), one can see that $u_i=0$. For the $n=2$
case, $u_2=0$ corresponded to choosing the uniformising coordinate (on
the upper half plane).  Similarly, here such a coordinate can be
thought of as choosing the uniformising coordinates.

Deformations of projective structure are obtained by deforming the
Halphen invariants by symmetric differentials. Hence the space of
projective structures can be identified by the space of symmetric
differentials just as in the case of $n=2$, a deformation of a
projective structure could be described by a quadratic differential.
This can be made more precise by using the correspondence with Higgs
bundle description of Teichm\"uller space\cite{hitchin,proposal}.

\section{Conclusion and outlook}

In this paper, we have constructed $W$-space using the generalised KdV
times as $w$-coordinates. This provides the geometry behind
$W$-gravity and also gives geometrical insight into the Teichm\"uller
components of Hitchin. The new set of complex manifolds
$W_n(\Sigma)$ that we have introduced may be interesting in their own
right. It would be interesting to obtain more information about them.
We believe that they must be Kahler manifolds like the original
manifold $\Sigma$. What is the metric on this space? It is also of
interest to understand the mapping class group (equivalent to
understanding ``large'' $W$-diffeomorphisms) so that we can construct
the moduli spaces for $W$-gravity. This will be of some significance for
$W$-strings.

Since the moduli of these spaces correspond to variation of Hodge
structures\cite{GLM,falqui}, there must exist a description of these
structures directly in the higher dimension. This can be done via
Higgs bundles in higher dimensions which has been studied by
Simpson\cite{simpson}.  Assuming this is possible, this would imply
that the differential equation (\ref{eodeb}) and the equations
(\ref{eexta}) can be rewritten as self-duality equations in the higher
dimensional space.  The proposal in \cite{proposal} of the flat
connections as generalised vielbeins and spin-connections can probably
be made more concrete.  This would lead to a covariant description of
$W$-gravity. We hope to discuss this issue in the future.

One of the important issues is to understand how to couple
matter to $W$-gravity directly in this higher dimensional setting.
Since objects in this higher dimensional space project down to terms
involving higher derivatives, it is obvious that extrinsic geometry
has to play a role. An interesting example has been furnished by the
case of a massless rigid particle whose action is given only by the
extrinsic curvature term. It has been shown recently that this theory
has $W_3$ symmetry\cite{wparticle}. There must exist stringy
generalisations of this with $W_n$ symmetry. This could possibly be
related to the QCD string or even new fundamental string theories.

This work has dealt with the case of Riemann surfaces with genus
$g>1$.  It is easy to include the cases of the sphere with $n\geq3$
punctures and the torus with $n>0$ punctures since they also admit
Fuchsian uniformisation. In the differential equation approach
studied in this paper, punctures correspond to regular singular
points of the differential equation such that the monodromy around
the puncture belong to a specific conjugacy class.
It is useful to work out this case since
one can be a lot more explicit and this can provide us with a better
understanding of the $w$-moduli.
Finally, even though we have
restricted our discussion to the case of $W_n$-gravity, the
generalisation to include the case of other simple Lie groups is
straight forward. This should cover the class of $W$-gravities which
are called ``generic''. It is not clear to us how the ``exotic''
algebras fit into this construction.

\appendix
\begin{flushleft}
  {\large\bf Appendix}
\end{flushleft}
\section{Wronskians}\label{wronk}
We discuss some elementary facts about linear differential equations.
Let $\{f_i\}$, $i=1,\ldots,n$ be a set of linearly independent
solutions of a $n$-th order homogeneous linear differential equation
of the form (\ref{eodea}). Given $\{f_i\}$, one can reconstruct the
differential equation (\ref{eodea}). This is done as follows. Define
the {\it Wronskian} as follows
\begin{equation}
W\equiv \left|\matrix{f_1 & f_2 & \cdots & f_n \cr
                 Df_1 & Df_2 & \cdots & Df_n \cr
                 \vdots & \vdots & & \vdots\cr
                 D^{n-1}f_1 & D^{n-1}f_2 & \cdots & D^{n-1}f_n\cr}\right|
\quad.  \label{wronskian}
\end{equation}
We shall refer to the Wronskian by $W[f_i]$ or just $W$ when its
arguments are obvious.  Further, define $W_i\equiv$ \{$W$ with the
$i$-th row replaced by $(D^{n}f_1,\ D^{n}f_2 ,\ \cdots ,\
D^{n}f_n)$\}. Then, $\{f_i\}$ solve equation (\ref{eodea}) with
$\hat{u}_i = -(W_i/W)$.  The proof of this is simple. The linear
independence of the solutions implies that any other solution $f$ will
be a linear combination of the $\{f_i\}$.  This can be expressed in
the following form from which the differential equation for $f$ is
obtained.
\begin{equation}
\tilde{W}\equiv \left|\matrix{f &f_1  & \cdots & f_n \cr
                 Df &Df_1 & \cdots & Df_n \cr
                 \vdots & \vdots & & \vdots\cr
                 D^n f &D^{n}f_1 &  \cdots & D^{n}f_n \cr}\right|
 = 0\quad.  \label{eodeaa}
\end{equation}
Comparing equation (\ref{eodeaa}) with (\ref{eodea}), we obtain
$\hat{u}_i = -(W_i/W)$.

One can check that under the substitution $f_i=\lambda(z) s_{i}$,
the Wronskian satisfies the following property
\begin{equation}
W[f_i] = \lambda^n W[s_{i}]\quad {\rm for} \quad i=1,\ldots,n\quad,
  \label{ewprop}
\end{equation}
for any function $\lambda(z)$.

One can easily see that $W_1= (-)^{n+1} DW$, which implies that
$\tilde{u}_1 = (-)^n D\ ln(W)$. So if $\tilde{u}_1=0$ (as in equation
(\ref{eodeb})), then it follows that $DW=0$ which implies that $W$ is
independent of $z$. From the definition of the Wronskian it follows
that $W\in K^{n(j+{{n-1}\over2})}$ if $f_i\in K^j$. Here $K$ is
the canonical holomorphic line bundle on a Riemann surface with
coordinate $z$. For the condition $W=$ constant to make sense globally,
we require that $W$ be a scalar ($W\in K^0$) which implies that $f_i
\in K^{{1-n}\over2}$.

\noindent {\bf Acknowledgements:} I would like to thank T. Jayaraman
for numerous discussions. Most of these ideas presented here arose out
of those discussions. I would also like to thank Pablo Ares Gastesi
for patiently explaining a lot of the mathematical ideas and finally,
I would like to thank Spenta Wadia for encouragement.


\begin{thebibliography}{99}
\bibitem{zam} A. B. Zamolodchikov, Teor. Mat. Fiz. {\bf 65} (1985)
1205.
\bibitem{DS} V. Drinfeld and V. Sokolov, J. Sov. Math. {\bf 30} (1984) 1975.
\bibitem{reviews}P. Bouwknegt and K. Schoutens, Phys. Rep. {\bf 223}
  (1993) 186;
L. Feher, L. O'Raifeartaigh, P. Ruelle, L. Tsutsui and A. Wipf, Phys.
Rep. {\bf 222} (1992) 1.
\bibitem{dbg} J. de Boer and J. Goeree, Nuc. Phys. {\bf B401} (1993)
369.
\bibitem{proposal} S. Govindarajan and T. Jayaraman, ``A proposal for
  the geometry of $W_n$-gravity,''
Phys.  Lett. {\bf B345}, 211 (1995)= hep-th/9405146.
\bibitem{hull} C. M. Hull, Commun. Math. Phys. {\bf 156} (1993) 245;
Nuc. Phys.{\bf B 413} (1994) 296.
\bibitem{hitchin} N. J. Hitchin, ``Lie groups and Teichm\"uller
  space,'' Topology {\bf 31} (1992) 451-487.
\bibitem{hejhal} D. A. Hejhal, ``Monodromy groups for higher-order
  differential equations,'' Bull.  AMS, {\bf 81} (1975) 590-592;
  ``Monodromy groups and linearly polymorphic functions,'' Acta
  Mathematica, {\bf 135} (1975) 1-55; ``Monodromy groups and
  Poincar\'e series,'' Bull. AMS {\bf 85} (1978) 339.
\bibitem{GLM} A. Gerasimov, A. Levin and A. Marshakov, ``On
  $W$-gravity in two dimensions,'' Nuc. Phys. {\bf B360} (1991)
  537-558.
\bibitem{gervais} J.-L.  Gervais and Y. Matsuo, ``Classical $A_n$ W -
  geometry,'' Commun.  Math. Phys. {\bf 152} (1993) 317-368; ``W
  geometries,'' Phys. Lett.  {\bf B274} (1992) 309-316.
\bibitem{gomis} J. Gomis, J. Herrero, K.  Kamimura and J. Roca,
  ``Finite $W_3$ transformations in a multi-time approach,''
  hep-th/9409024.
\bibitem{sotkov} G. Sotkov and M. Stanishkov, ``Affine geometry and
  $W_n$ gravities,'' Nuc. Phys. {\bf B356} (1991) 439-468.
\bibitem{itzykson} C. Itzykson, ``$W$ - geometry,'' Cargese lectures,
  appeared in {\it Random surfaces and Quantum gravity}, ed. O.
  Alvarez {\it et al.}, Plenum press, New York, 1991.
\bibitem{alvarezgaume} L. Alvarez-Gaum\'e and C. Gomez, ``Topics in
  Liouville theory,'' Lectures at the Spring School on String Theory
  and Quantum Gravity, Trieste, Italy, Apr 15-23, 1991.  Published in
  Trieste Spring School 1991:142-177.
\bibitem{falqui} E. Aldrovandi and G. Falqui, ``Geometry of Higgs and
  Toda Fields on Riemann Surfaces,'' Preprint hep-th/9312093; See also
  hep-th/9411184.
\bibitem{diFr} P. Di Francesco, C. Itzykson and J.-B. Zuber,
  ``Classical $W$ -- algebras,'' Comm. Math. Phys. {\bf 140} 543-567.
\bibitem{hawley} N. Hawley and M. Schiffer, ``Half-order differentials
  on Riemann surfaces,'' Acta. Math. {\bf 115} (1966) 199-236.
\bibitem{forsyth} A. R. Forsyth, {\it Theory of differential
    equations}, Vol. 4, Dover Publications, New York.
\bibitem{wilc} E. J. Wilczynski, {\it Projective differential geometry
    of curves and ruled surfaces}, Chelsea Publishing company, New
  York (1905).
\bibitem{jimbo}
  M. Jimbo and T. Miwa, ``Monodromy preserving deformations of linear
  ordinary differential equations with rational coefficients I and
  II,'' Physica {\bf 2D} (1981) 306; Physica {\bf 2D} (1981) 407.
\bibitem{its} A. Its and V. Yu. Novokshenov, {\it The isomonodromic
    deformation method in the theory of Painlev\'e equations}, Lec.
  Notes in Mathematics, vol 1191, Springer.
\bibitem{novikov} S. P. Novikov, ``The periodic problem for the
  Korteweg-de Vries equation,'' Func. Anal. and Appl. {\bf 8} (1974)
  236-246.
\bibitem{matsuo} Y.  Matsuo, ``Classical $W_n$ symmetry and the
  grassmanian manifold,'' Phys. Lett. {\bf B277} (1992) 95-101.
\bibitem{kobayashi} S. Kobayashi, {\it Hyperbolic manifolds and
    holomorphic mappings}, Marcel Dekker, New York, 1970.
\bibitem{zucchini} R. Zucchini, ``Light cone $W_n$ geometry and its
  symmetries and projective field theory,'' Class. Quant. Grav. {\bf
    10} (1993) 253-278.
\bibitem{flat} A. Bilal, V. V. Fock and I. I. Kogan, Nuc. Phys. {\bf
    B359} (1991) 635; A. Das, W. -J. Huang and S. Roy, Int. J. of Mod.
  Phys. {\bf A7} (1992) 3447; J de Boer and J. Goeree, Nuc. Phys. {\bf
    B381} (1992) 329-359; Phys. Lett. {\bf B274} (1992) 289-297; K.
  Yoshida, Int. J. of Mod. Phys. {\bf A7} (1992) 4353-4376.
\bibitem{beltrami} S. Govindarajan, ``Covariantising the Beltrami equation in
   W - geometry," Preprint hep-th/9504003.
\bibitem{simpson} C. Simpson, ``Constructing variations of Hodge
  structure using Yang-Mills theory and applications to
  uniformization,'' J. of the AMS {\bf 1} (1988) 867-918.
\bibitem{wparticle} E. Ramos and J. Roca, ``$W$-symmetry and the rigid
  particle,'' Preprint hep-th/9408019.
\end{thebibliography}
\end{document}